\documentclass[12pt]{article}
\begin{document}
\pagestyle{plain}
\setcounter{page}{1}
%\preprint{UTPT-99-02}
\begin{center}
{\large\bf Finite Higher-Dimensional Unified Field Theory and TeV Physics}
\vskip 0.3 true in
{\large J. W. Moffat}
%\date{ }
\vskip 0.3 true in
{\it Department of Physics, University of Toronto,
Toronto, Ontario M5S 1A7, Canada}
\end{center}
\date{\today}

\begin{abstract}%
A unified field theory based on the compactification of a higher 
D-dimensional Einstein-Yang-Mills-Higgs action is developed. 
The extra $D-4$ dimensions form a compact internal space with scale size 
$R$. An anomaly-free unified chiral model of quarks and leptons, described 
by $SO(18)$ in twelve dimensions, breaks down to $SO(10)\times 
SO(8)\rightarrow SO(10)$ with a non-trivial topological structure and three
chiral families in four dimensions. A quantum field theory formalism in 
$D$-dimensions leads to a self-consistent, finite quantum gravity, 
Yang-Mills and Higgs theory, which is unitary and gauge invariant to all 
orders of perturbation theory. The gauge hierarchy problem is solved due to
the exponential damping of the Higgs self-energy loop graph for energies 
greater than $\sim 1$ TeV, and because of the reduction of quantum gravity 
to a scale of several TeV. The compactification scale is $M_c\geq 1$ TeV, 
leading to Kaluza-Klein excitations and experimental signatures at a scale 
of several TeV. Various scenarios for evading fast proton decay are 
discussed. \end{abstract}

%\vskip 0.2 true in

%\pacs{ }

\section{Introduction}

The standard model has been verified to remarkable accuracy\cite{Altarelli}
down to scales of $10^{-15}$ cm, corresponding to energies up to $\simeq 
100$ GeV. With the discovery of the top quark with a mass $m_t=175.6(5.5)$
GeV, all the required fermions in the standard model are now in place. 
The Higgs particle, which represents a vital missing element in the 
standard model, is yet to be discovered. When it is found, we could just 
declare that particle physics is closed. However, there are conceptual 
difficulties with the standard model, which point to new physics beyond it.
A successful unified theory of gravity and the standard model should at 
least accomplish the following:

\begin{enumerate}

\item 

Resolve the gauge hierarchy problem. The gauge hierarchy (or 't Hooft 
naturalness) problem besets the Higgs sector. The standard model 
cannot naturally explain the relative smallness of the weak scale of mass, 
set by the Higgs mechanism at $M_{\rm WS}\sim 250$ GeV.

\item  

Reduce the number of unknown parameters.

\item 

Explain the origin of the three fermion generations in the standard model.

\item 

Provide a mathematically consistent quantum gravity theory which leads to
finite scattering amplitudes to all orders in perturbation theory. 

\item

Guarantee that the proton remains stable in accordance with the current 
experimental lower bound on its decay lifetime $\tau_p$.

\end{enumerate} 

A leading candidate for a unified theory of the standard model and gravity
has been superstring theory\cite{Schwarz}. String theory 
bypasses the problem of ultraviolet divergences of gravity by replacing 
the fundamental point-like object by a string, an extended one-dimensional 
object. String compactifications lead to a myriad of possible vacuum states
as models of low energy particle physics, but recent developments in 
duality and p-branes\cite{Duff} have led to new possibilities for 
unification models. D-branes are associated with gauge fields living in 
their world volume\cite{Witten,Dienes,Shiu}. The standard model gauge group
would correspond to gauge fields living in the world volume of 3-branes. If
the gauge group comes from open strings starting and ending on a set of 
p-branes, then the string scale $M_s$ can be lowered much below the Planck 
mass scale, $M_{\rm Planck}\sim 10^{19}$ GeV, by using the formula (for 
3-branes): $M^4_s=\alpha_{\rm GUT}M_c^3 M_{\rm Planck}/\sqrt{2}$, where 
$M_c$ is the compactification scale. If $M_s\sim 1$ TeV, then predictions 
could be made that might be checked by the new generation of accelerators. 
Dienes et al.\cite{Dienes2} have shown that a Kaluza-Klein orbifold 
reduction can lead to a gauge coupling unification at a much lower energy 
scale than the usual GUT scale with power law behaviour instead of the 
familar logarithmic scaling behaviour. A more radical idea has been to 
reduce the Planck scale of gravity to the TeV energy region\cite{Arkani}, 
thus resolving the hierarchy problem.

In previous work\cite{Moff}, the author developed a model based on the 
gauge group ${\cal G}=SO(3,1)\times SU_c(3)\times SU(2)\times U(1)$ in four
dimensions. No attempt was made to unify the standard model with 
gravity. The unknown parameters such as coupling constants and fermion 
masses were to be determined by a relativistic, Schr\"odinger-type 
eigenvalue equation, using the perturbatively finite and unitary formalism 
of finite quantum field theory (FQFT)[10-18]. The stability of the proton 
was guaranteed, for quarks and leptons were not combined in the same 
irreducible fermion representation. 

If, indeed, baryon number is conserved, then the small but non-zero matter 
content of the universe is simply a matter of the initial conditions, and 
its value cannot be explained within the standard domain of physics, which 
is not a satisfactory state of affairs. Moreover, it is difficult to find 
non-trivial solutions to the relativistic eigenvalue equation for the 
mass spectrum and the coupling constants. In view of this, it is tempting 
to pursue further the possibility of discovering a unified field theory of 
gravity and the gauge fields of the standard model.

In the following, we shall pursue such a possible theory based on a 
higher-dimensional unified field theory. In the early 
eighties there was a revival of attempts to build a unified theory using 
the idea of a Kaluza-Klein pure gravity theory in 
D-dimensions\cite{Kaluza}. These ideas were abandoned when it was 
discovered that assuming a Riemannian geometry in D-dimensions and an 
internal compact space, the models failed to predict flavor chiral fermions
in the four-dimensional theory\cite{Wetterich}. Beginning with a spinor 
coupled to gravity in D dimensions, one always ends in the four-dimensional
theory with vector-like fermion representations of the gauge group. 
Applying a theorem due to Atiyah and Hirzebruch\cite{Hirzebruch}, it is 
found that the number of chiral fermions derived from dimensional reduction
of a Weyl spinor coupled to Riemannian geometry with $D-4$ dimensions 
describing a compact, orientable manifold without boundary is zero. This 
situation is also found to exist for $N=8$ supergravity theories.
Moreover, except for certain special choices of parameters, the fermions 
have masses of order the Planck mass. An explanation of the observed small 
fermion masses requires chiral fermions, where the left-handed and 
right-handed Weyl or Weyl-Majorana spinors belong to inequivalent 
representations of the low energy group. Another reason for abandoning the 
Kaluza-Klein approach to unification is that the gravity theory is not 
renormalizable. The usual ultraviolet divergences that plague quantum 
gravity are present, as in the standard point particle four-dimensional 
quantum gravity theory.

We shall base our unification theory on a D-dimensional Einstein-Yang-
Mills-Higgs field theory, with supplementary gauge fields coupled to 
gravity, which has chiral fermion representations corresponding
to massless Dirac modes, and is free of anomalies. We consider an 
$SO(18)\supset SO(10)\times SO(8)$ model in twelve dimensions in which the 
eight-dimensional internal space is described by spinor connection gauge 
fields, associated with an $SO(8)$, which are topologically non-trivial. By
dimensional reduction the $SO(18)$ leads to the four-dimensional grand 
unified theory (GUT) $SO(10)$ and the $SO(8)\supset Sp(4)\times SU(2)$ with
three families of chiral quarks and leptons. 

The FQFT gauge formalism is applied to the D-dimensional theory to 
guarantee a self-consistent quantum gravity theory coupled to the 
Yang-Mills, Higgs and spinor fields. The formalism is free of tachyons and 
unphysical ghosts and satisfies unitarity to all orders of perturbation 
theory. It could incorporate supersymmetry if required, in the form of a 
supergravity theory, but we shall not do so here, in order to aim for as 
minimal a scheme as possible. The gauge hierarchy problem is resolved 
because the finite scalar Higgs self-energy loop graphs are damped 
exponentially at high energies above the physical Higgs scale $\Lambda_H$ 
set by the FQFT formalism and by choosing $\Lambda_H\sim 1$ TeV. The 
compactification scale $M_c$ can be as low as $M_c\simeq 1-10$ TeV, while 
the quantum gravity scale $\Lambda_G\sim 1-10$ TeV. This would predict that
at these energies future high-energy experiments could detect Kaluza-Klein 
excitation modes at energies of several TeV.

If we choose $\Lambda_G\sim 1-10$ TeV, then quantum gravity loop 
corrections are perturbatively weak all the way to the Planck energy. This 
would obviate the need to find a non-perturbative quantum gravity 
formalism.

\section{\bf Kaluza-Klein Theory and The Ground State}

We shall begin with the action:
\begin{equation}
W=W_{\rm grav}+W_{YM}+W_{\rm H}+W_{\rm Dirac},
\end{equation}
where
\begin{equation}
W_{\rm grav}=-\frac{1}{\kappa^2}\int d^Dz\sqrt{-g}(R+\lambda),
\end{equation}
\begin{equation}
W_{\rm YM}=-\frac{1}{4}\int d^Dz\sqrt{-g}\,{\rm tr}(F^2),
\end{equation}
\begin{equation}
W_{\rm H}=-\frac{1}{2}\int d^Dz\sqrt{-g}[D_M\phi^aD^M\phi^a+V(\phi^2)], 
\end{equation}
\begin{equation}
W_{\rm Dirac}=\frac{1}{2}\int 
d^Dz\sqrt{-g}\bar{\psi}\Gamma^Ae_A^M[\partial_M\psi-\omega_M\psi -{\cal 
D}(A_M)\psi]+h.c. \end{equation} 
Here, we use the notation: $z^M=(x^\mu; 
\mu=0,1,2,3, y^m; m=1,2,...,D-4), M=0,...,D, g={\rm det}(g_{MN})$. The 
Riemann tensor is defined such that 
${{R_{LM}}^K}_N=\partial_L{\Gamma_{MN}}^K-\partial_M{\Gamma_{LN}}^K+
{\Gamma_{LC}}^K{\Gamma_{MN}}^C-{\Gamma_{MC}}^K{\Gamma_{LN}}^C$.
Moreover, h.c. denotes the Hermitian conjugate, $\bar\psi=\psi^{\dagger}
\Gamma^0$, and $e^M_A$ is a vielbein, related to the metric by
\begin{equation}
g_{MN}=\eta_{AB}e_M^Ae_N^B, 
\end{equation}
where $\eta_{AB}$ is the D-dimensional Minkowski metric tensor associated
with the flat tangent space with indices A,B,C... Moreover, $F^2=
F_{MN}F^{MN}$, $R$ denotes the scalar curvature, $\lambda$ is
the cosmological constant and
\begin{equation}
F_{aMN}=\partial_NA_{aM}-\partial_MA_{aN}-ef_{abc}A_{bM}A_{cN},
\end{equation}
where $A_{aM}$ are the gauge fields of the Yang-Mills group with
generators $f_{abc}$ and $e$ is the coupling constant. $\kappa^2=16\pi
\bar G $, where $\bar G$ is related to Newton's constant $G$ by $\bar 
G=GV$ and $V$ is the volume of the internal 
space. The dimensions of $\bar G$ are $(\rm length)^{\delta}$($\delta=D-2$). Moreover, 
$D_M$ is the covariant derivative operator:
\begin{equation}
D_M\phi^a=\partial_M\phi^a+ef^{abc}A_M^b\phi^c.
\end{equation}
The Higgs potential $V(\phi^2)$ is of the form leading to spontaneous 
symmetry breaking
\begin{equation}
V(\phi^2)=\frac{1}{4}g(\phi^a\phi^a-K^2)^2+V_0,
\end{equation}
where $V_0$ is an adjustable constant and the coupling constant $g > 0$.

The spinor field is minimally coupled to the gauge potential $A_M$, and 
${\cal D}$ is a matrix representation of the gauge group $G$ defined in
D-dimensions. The spin connection $\omega_M$ is
\begin{equation}
\omega_M=\frac{1}{2}\omega_{MAB}\Sigma^{AB},
\end{equation}
where $\Sigma^{AB}=\frac{1}{4}[\Gamma^A,\Gamma^B]$ is the spinor matrix 
associated with the Lorentz algebra $SO(D-1,1)$. The components 
$\omega_{MAB}$ satisfy
\begin{equation}
\partial_Me^K_A+{\Gamma_{MN}}^Ke^N_A-{\omega_{MA}}^Be_B^K=0,
\end{equation}
where ${\Gamma_{MN}}^K$ is the Christoffel symbol.  
    
The field equations for the gravity-Yang-Mills-Higgs-Dirac sector are
\begin{equation}
R_{MN}-\frac{1}{2}g_{MN}R=-\frac{1}{2}\kappa^2(T_{MN}-\lambda g_{MN}),
\end{equation}
\begin{equation}
g^{LM}\nabla_LF_{MN}=g^{LM}\biggl(\partial_LF_{MN}-{\Gamma_{LM}}^KF_{KN}
-{\Gamma_{LN}}^KF_{MK}
$$
$$
+[A_L,F_{MN}]\biggr)=0,
\end{equation}
\begin{equation}
\frac{1}{\sqrt{-g}}D_M[\sqrt{-g}g^{MN}D_N\phi^a]
=\biggl(\frac{\partial V}{\partial\phi^2}\biggr)\phi^a,
\end{equation}
\begin{equation}
\Gamma^Ae^M_A[\partial_M-\omega_M-{\cal D}(A_M)]\psi=0.
\end{equation}
The Yang-Mills-Higgs contribution to the energy-momentum tensor is
\begin{equation}
T_{MN}^{\rm YMH}={\rm tr}(F_{MK}F^K_N)+D_M\phi^aD_N\phi^a
-\frac{1}{2}g_{MN}\biggl[\frac{1}{2}{\rm tr}(F^2) 
$$
$$
+D_P\phi^aD^P\phi^a+V(\phi^2)\biggr].
\end{equation}

We must now choose an ansatz for the ground state of the four-dimensional 
world. A general theory would start by assuming the ground state to be 
$M^4\times B$, where $M^4$ is four-dimensional Minkowski space and $B$ is 
a compact internal space. A simple ansatz for the compact space $B$ is 
to assume a symmetric solution with the structure, $M^4\times S/H$, where
$S/H$ is a coset space of dimension $D-4$. For the metric we take 
\begin{equation} 
g_{MN}dz^Mdz^N=g_{\mu\nu}(x)dx^\mu dx^\nu+g_{mn}(y)dy^m 
dy^n, 
\end{equation} 
\begin{equation}
A^a_Mdz^M=A^a_\mu dy^\mu.
\end{equation}
One possible symmetric choice for the ground state four-dimensional 
spacetime is that $g_{\mu\nu}$ is a de Sitter solution of the 
four-dimensional Einstein equation: \begin{equation}
R_{\mu\nu}=-\frac{1}{2}\kappa^2\lambda g_{\mu\nu}.
\end{equation}
Since we have additional Yang-Mills and Higgs scalar fields in our 
higher-dimensional theory, it is possible for us to obtain classical 
solutions to the field equations in spacetime as the product of flat 
four-dimensional Minkowski spacetime and an internal compact space. This 
"spontaneous compactification" can be achieved by going beyond a pure 
Kaluza-Klein theory, which does not allow a flat four-spacetime unless the 
curvature of the internal space is also zero\cite{Cremmer,Luciani}. 

If the metric $g_{MN}$ for the space $M^4\times B$ is by construction a 
solution of the Killing equation for $M^4\times B$, then one can carry out 
the integrations over the $y$ coordinates in the action, and the dynamical 
variables in the theory are functions of $x$ only. The dimensional 
reduction reduces some of the gauge fields $A_M$ to $A_\mu(x)$, while 
certain linear combinations of the gauge fields $A_m(x,y)$ become 
geometrical scalar fields in four dimensions. However, the latter scalar 
fields do not in general lead to a spontaneous symmetry breaking 
Higgs mechanism, so that our additional Higgs field action $W_H$
is required to perform this task.

\section{\bf Reduction to a Flavor-Chiral Theory}

The problem of obtaining the correct quark and lepton quantum numbers is 
more subtle and difficult than one might suspect in both Kaluza-Klein 
theories and in string theories. One of the most striking features of 
particle physics is the knowledge that the quantum numbers of fermions 
{\it are not vector-like}, i.e., that left-handed fermions 
transform under $SU(3)\times SU(2)\times U(1)$ differently from the way 
right-handed fermions transform. Left-handed quarks are $SU(2)$ doublets 
but right-handed quarks are  $SU(2)$ singlets. Fermions of given helicity 
form a complex representation of $SU(3)\times SU(2)\times U(1)$, so the 
fermion representations are not self-conjugate. This fact plays an 
important role in unified theories, because it means that the bare masses 
of the quarks and leptons are ruled out by gauge invariance. The fermions 
can acquire mass only through spontaneous symmetry breaking. Thus, arises 
the problem of explaining the relative lightness of observed fermions. What
generates the smallness of the $SU(2)\times U(1)$ breaking scale? 

Since the quantum numbers of the fermions are not vector-like, the spectrum
of light fermions depends only on universality class features of an 
$SU(3)\times SU(2)\times U(1)$ invariant theory, i.e. the lightness of the 
fermions cannot be modified by any $SU(3)\times SU(2)\times U(1)$ 
invariant perturbations. We shall consider for the present that the light 
fermions are massless, ignoring the $SU(2)\times U(1)$ breaking. 

Another striking feature of the fermion spectrum is that the anomalous 
triangle graphs cancel, an important ingredient in a successful gauge 
invariant unified theory. In addition, each family of fermions consists of 
five irreducible representations of $SU(3)\times SU(2)\times U(1)$ and 
their exists a redundancy of three families. The $SU(5)$ and $SO(10)$ grand
unified models successfully describe the fermion structure in one family 
in terms of the representation 
$\underline{\overline{5}}_L+\underline{10}_L$ in $SU(5)$, and 
$\underline{16}_L$ in $SO(10)$\cite{Glashow,Fritzsch}. As far as the
replication of families is concerned, it seems natural and elegant to 
describe this replication in terms of spinor representations of $SO(N)$ for
$N\geq 18$. 

Witten\cite{Witten2} proved by using topological arguments that, in any 
number of dimensions, the Dirac operator in a pure Kaluza-Klein 
theory cannot admit a chiral spectrum. Wetterich\cite{Wetterich} showed 
that the spectrum of fermions could be chiral only if the dimensionality of
the space is 2 mod 8. Only if additional Yang-Mills gauge fields are 
included in a higher-dimensional theory can the fermion spectrum become 
chiral under dimensional reduction. However, this is only true if the 
compactification involves a topologically non-trivial configuration of 
these gauge fields. The addition of extra gauge symmetries is indeed the 
mechanism whereby superstring theories lead to a spectrum of chiral 
fermions.

In higher-dimensional theories with a ground state $M^4\times B$, a 
massless spectrum of particles is generated as zero modes of wave operators
on the internal compact space $B$. A massless fermion particle in $D=4+n$ 
dimensions obeys
\begin{equation}
\label{Dirac}
\Gamma^MD_M\psi=0,
\end{equation}
where $\Gamma^M$ are the gamma matrices. The quantum numbers remain 
unchanged in the presence of non-minimal couplings, so we ignore them in 
the present discussion. We can separate (\ref{Dirac}) into
\[
D^{(4)}\psi+D^{(n)}\psi=0,
\]
where $D^{(4)}=\Gamma^\mu D_\mu$ and $D^{(n)}=\Gamma^mD_m$, and we see that
$D^{(n)}$ is a mass operator, whose eigenvalues are the experimentally 
determined fermion masses. The zero eigenvalues are the massless fermions.

In order to see that a pure Riemannian Kaluza-Klein theory 
has difficulties, we employ an argument due to 
Lichnerowicz\cite{Lichnerowicz}. If we square the internal Dirac operator, 
we get $(iD^{(n)})^2=-D_mD^m+\frac{1}{4}R$,
and because $-D^mD_m$ is a non-negative operator, then if $R>0$ 
everywhere, it follows that the Dirac operator has no zero eigenvalues, 
i.e. there are no massless fermions. By using the Atiyah-Hirzebruch 
theorem\cite{Hirzebruch}, which states that the character-valued index of 
the Dirac operator vanishes on any manifold with a continuous symmetry 
group (in any even number of dimensions), Witten proved in general that a 
compact Riemannian manifold (such as the internal space of a pure 
Kaluza-Klein theory) does not possess zero mass chiral fermions.

We shall now assume that we have additional Yang-Mills gauge fields. If the
gauge quantum numbers of the fermions are vector-like, they will remain 
vector-like after compactification, unless special Majorana and 
Weyl-Majorana conditions are imposed together with gauge symmetry 
conditions\cite{Slansky}. A serious problem of cancellation of 
anomalies will occur, unless careful attention is payed to the gauge 
couplings and the compactification. Non-vector-like couplings of gauge 
fields will lead to anomalies, unless the condition is 
satisfied\cite{Witten2}: 
\begin{equation}
\label{anomaly}
{\rm tr}(M_L^a)^r={\rm tr}(N_R^a)^r,
\end{equation}
where $M_L^a$ and $N_R^a$ are matrices that couple the gauge fields $A_m^a$
to left-handed and right-handed spin $1/2$ fermions, respectively. 
Moreover, $r=n+1, n-1, n-3,n-5,...$ in $2n$ dimensions with $n+1$ external 
gluons, $n-1$ external gluons and two gravitons, $n-3$ external gluons and 
four gravitons, etc. which all correspond to anomalous graphs. 

Witten\cite{Witten2} found solutions 
to Eq.(\ref{anomaly}) in $2n$ dimensions, which we shall use in the 
following. Let us consider in $2n$ dimensions, a theory with the orthogonal
gauge group $SO(2n+6)$. We assert that the positive chirality spinors of 
the Lorentz group $SO(2n-1,1)$ transform as positive chirality spinors of 
the gauge group $SO(2n+6)$, while the negative chirality $SO(2n-1,1)$ 
spinors transform as negative chirality spinors of $SO(2n+6)$. For any $n 
> 0$ this leads to an anomaly-free theory, and for $n=2$, it is the 
familiar $SO(10)$ grand unified model in four dimensions\cite{Fritzsch}.
 
We must guarantee that, if we begin with a non-vector-like theory in $2n$ 
dimensions, then we retain this property under dimensional reduction. The 
condition
\[
\Gamma^{(4+n)}=\Gamma^{(4)}\cdot\Gamma^{(n)}
\]
where $\Gamma^{(4+n)}=\Gamma_1...\Gamma_{4+n}, 
\Gamma^{4}=\Gamma_1...\Gamma_4$, and 
$\Gamma^{(n)}=\Gamma_5...\Gamma_{4+n}$, shows that the $4+n$-dimensional 
chirality operator $\Gamma^{(4+n)}$ differs from $\Gamma^{(4)}$ by a factor
$\Gamma^{(n)}$ that can equal plus 1 or minus 1, so that we can lose the 
non-vector-like property under compactification. As shown by 
Randjibar-Daemi, Salam and Strathdee\cite{Salam}, this can be avoided by 
attributing a non-trivial topological structure to the internal gauge 
group ${\cal K}$, associated with the internal Kaluza-Klein space. In 
particular, this can be achieved by inserting a generalized Dirac monopole 
inside the Kaluza-Klein space.

Let us consider an $SO(18)$ theory in twelve dimensions. We identify the
internal space spinor connections $\omega^a_j$ with gauge fields $B^a_j$, 
which have a non-vanishing vacuum expectation value, $\langle 
B^a_j\rangle_0\not= 0$. The $B^a_j$ are $SO(8)$ gauge fields on the 
eight-dimensional Riemannian manifold of the Kaluza-Klein sector. The 
$SO(8)$ is embedded in $SO(18)$ , such that we obtain the symmetry breaking
$SO(18)\rightarrow SO(8)\times SO(10)$, which leads to the breaking in four
dimensions: $SO(8)\times SO(10)\rightarrow SO(10)$. Depending on the type 
of manifold assumed for the compact dimensions, we get different 
numbers of families which equal the Euler characteristic of the space $B$. 
The number of families in ten dimensions is always even. For example, for 
the manifolds $S^2\times S^4$ and $CP^3$ the number of families of zero 
mass fermions is four, while for $S^2\times S^2\times S^2$ the number of 
fermion families is eight. For models based on eight and twelve dimensions,
the number of fermion families is odd.

Since the observed number of chiral families in four dimensions is 
three, we shall restrict our attention to the orthogonal group $SO(18)$ in 
twelve dimensions with an odd number of families. We can decompose the 
$\underline{256}$-dimensional representation of $SO(18)$ as 
\[
(\underline{8}_{sp},\underline{\overline{16}})+(\underline{8}'_{sp},\underline{16})
\]
of $SO(8)\times SO(10)$, where $\underline{8}_{sp}$ and 
$\underline{8}'_{sp}$ are the two real inequivalent spinors of $SO(8)$. Now
consider $SO(8)\supset Sp(4)\times SU(2)$, where the vectorial octet 
$\underline{8}_V$ of $SO(8)$ yields $(\underline{4},\underline{2})$ of 
$Sp(4)\times SU(2)$ and $\underline{8}'_{sp}$ of $SO(8)$ equivalently, 
whereas $\underline{8}_{sp}$ of $SO(8)$ yields 
\[
(\underline{1},\underline{3})+(\underline{5},\underline{1})
\]
of $Sp(4)\times SU(2)$. This then gives for the $\underline{256}$ 
representation of $SO(18)$:
\[
(\underline{1},\underline{3},\underline{16})+(\underline{5},\underline{1},\underline{16})
+(\underline{4},\underline{2},\underline{\overline{16}})
\]
of $Sp(4)\times SU(2)\times SO(10)$. By identifiying $Sp(4)$ as a 
supplementary factor of the exactly conserved colour group $SU^c(3)\times 
Sp^c(4)'$, and $SU(2)$ as a gauged family subgroup of $SO(18)$, then only 
the fundamental left-handed fermions without primed colour are three 
families of the sixteen-dimensional representation of $SO(10)$, which 
agrees with observations\cite{Gell-Mann,Zee}.

\section{\bf Finite Quantum Field Theory Formalism in D Dimensions}

It is well-known that higher-dimensional field theories are 
non-renormalizable, for their structure has enhanced divergences. The 
non-renormalizability arises from the presence of infinite towers of 
non-chiral Kaluza-Klein states which circulate in all Feynman graphs. Even 
if we choose to describe the space as consisting of D-dimensional flat 
spacetime, the field theory would still be non-renormalizable, because we 
need to integrate over D dimensions of uncompactified loop momenta. 
Therefore, the well-known non-renormalizability of quantum gravity is 
further exacerbated in higher-dimensional theories. However, by applying 
our finite quantum field theory (FQFT) formalism, based on a nonlocal interaction 
Lagrangian which is perturbatively finite, unitary and gauge invariant 
[10-18], we can obtain a finite quantum field theory in higher dimensions 
and, in contrast to string theory, achieve a {\it genuine quantum field 
theory}, which allows vertex operators to be taken off the mass shell.  
The finiteness draws from the fact that factors of $\exp[{\cal 
K}(p^2)/2\Lambda^2]$ are attached to propagators which suppress any 
ultraviolet divergences in Euclidean momentum space, where $\Lambda$ is an 
energy scale factor. An important feature of FQFT is {\it that only the 
quantum loop graphs have non-local properties}; the classical tree graph 
theory retains full causal and local behaviour.

An important development in FQFT was the discovery that gauge invariance and
unitarity can be restored by adding series of higher interactions. The
resulting theory possesses a nonlinear, field representation dependent 
gauge invariance which agrees with the original local symmetry on shell but
is larger off shell. Quantization is performed in the functional
formalism using an analytic and convergent measure factor which retains 
invariance under the new symmetry. An explicit calculation was made of the 
measure factor in QED\protect\cite{Moffat2}, and it was obtained to lowest 
order in Yang-Mills theory\protect\cite{Kleppe2}. Kleppe and 
Woodard\protect\cite{Woodard2} obtained an ansatz based on the derived 
dimensionally regulated result when $\Lambda\rightarrow\infty$, which was
conjectured to lead to a general functional measure factor in FQFT gauge 
theories.

A convenient formalism which makes the FQFT construction transparent is 
based on shadow fields\protect\cite{Kleppe2,Woodard2}. We shall consider 
the D-dimensional spacetime to be approximately flat Minkowski spacetime, 
which is a valid approximation for circles in the internal space $B$ having
large fixed radii $R$.

Let us denote by $f_i$ a generic local field and write the standard local 
action as 
\begin{equation}
W[f]=W_F[f]+W_I[f],
\end{equation}
where $W_F$ and $W_I$ denote the free part and the interaction part
of the action, respectively, and
\begin{equation}
W_F=\frac{1}{2}\int d^Dzf_i{\cal K}_{ij}f_j.
\end{equation}
In a gauge theory $W$ would be the Becchi, Rouet, Stora, Tyutin (BRST) 
gauge-fixed action including ghost fields in the invariant action required 
to fix the gauge\cite{Becchi}. The kinetic operator ${\cal K}$ is fixed by 
defining a Lorentz-invariant distribution operator: \begin{equation} {\cal 
E}\equiv \exp\biggl(\frac{{\cal K}}{2\Lambda^2}\biggr) \end{equation} and
the shadow operator: \begin{equation}
{\cal O}^{-1}=\frac{{\cal K}}{{\cal E}^2-1}.
\end{equation}

Every local field $f_i$ has an auxiliary counterpart field $h_i$, and they 
are used to form a new action:
\begin{equation}
W[f,h]\equiv W_F[\hat f]-P[h]+W_I[f+h],
\end{equation}
where
\[
\hat f={\cal E}^{-1}f,\quad P[h]=\frac{1}{2}\int d^Dzh_i{\cal
O}^{-1}_{ij} h_j.
\]
By iterating the equation
\begin{equation}
h_i={\cal O}_{ij}\frac{\delta W_I[f+h]}{\delta h_j}
\end{equation}
the shadow fields can be determined as functions, and the regulated 
action is derived from
\begin{equation}
\hat W[f]=W[f,h(f)].
\end{equation}
We recover the original local action when we take the limit 
$\Lambda\rightarrow\infty$ and $\hat f\rightarrow f, h(f)\rightarrow 0$.

Quantization is performed using the definition
\begin{equation}
\langle 0\vert T^*(O[f])\vert 0\rangle_{\cal E}=\int[Df]\mu[f]({\rm gauge\, 
fixing})O[\hat f]\exp(i\hat W[f]).
\end{equation}
On the left-hand side we have the regulated vacuum expectation value of the
$T^*$-ordered product of an arbitrary operator $O[f]$ formed from the local 
fields $f_i$. The subscript ${\cal E}$ signifies that a regulating Lorentz 
distribution has been used. Moreover, $\mu[f]$ is a measure factor and 
there is a gauge fixing factor, both of which are needed to maintain 
perturbative unitarity in gauge theories.

The new Feynman rules for FQFT are obtained as follows: The vertices remain 
unchanged but every leg of a diagram is connected either to a regularized 
propagator,
\begin{equation}
\label{regpropagator}
\frac{i{\cal E}^2}{{\cal K}+i\epsilon}
=-i\int^{\infty}_1\frac{d\tau}{\Lambda^2}\exp\biggl(\tau
\frac{{\cal K}}{\Lambda^2}\biggr),
\end{equation}
or to a shadow propagator,
\begin{equation}
-i{\cal O}=\frac{i(1-{\cal E}^2)}{{\cal K}}=-i\int^1_0\frac{d\tau}
{\Lambda^2}
\exp\biggl(\tau\frac{{\cal K}}{\Lambda^2}\biggr).
\end{equation}
The formalism is set up in Minkowski spacetime and loop integrals are 
formally defined in Euclidean space by performing a Wick rotation. This 
facilitates the analytic continuation; the whole formalism could from the 
outset be developed in Euclidean space.

In FQFT renormalization is carried out as in any other field theory. 
The bare parameters are calculated from the renormalized ones and 
$\Lambda$, such that the limit $\Lambda\rightarrow\infty$ is finite for
all noncoincident Green's functions, and the bare parameters are those of 
the local theory. The regularizing interactions {\it are determined by the 
local operators.}

The regulating Lorentz distribution function ${\cal E}$ must be chosen to 
perform an explicit calculation in perturbation theory. We do not know the 
unique choice of ${\cal E}$\cite{Feynman}. It maybe that there exists an 
equivalence mapping between all the possible distribution functions ${\cal 
E}$. However, once a choice for the function is made, then the
theory and the perturbative calculations are uniquely fixed. A standard
choice in early FQFT papers is\protect\cite{Moffat,Moffat2}:
\begin{equation}
\label{reg}
{\cal E}_m=\exp\biggl(\frac{\partial^2-m^2}{2\Lambda^2}\biggr).
\end{equation}

An explicit construction for QED was given using the Cutkosky rules as 
applied to FQFT whose propagators have poles only where ${\cal K}=0$ and 
whose vertices are entire functions of ${\cal K}$. The regulated action 
$\hat W[f]$ satisfies these requirements which guarantees unitarity on the 
physical space of states. The local action is gauge fixed and then a
regularization is performed on the BRST theory.

The infinitesimal transformation
\begin{equation}
\delta f_i=T_i(f)
\end{equation}
generates a symmetry of $W$, and the infinitesimal transformation
\begin{equation}
\hat\delta f_i={\cal E}^2_{ij}T_j(f+h[f])
\end{equation}
generates a symmetry of the regulated action ${\hat W}$. It follows 
that FQFT regularization preserves all continuous symmetries including 
supersymmetry. The quantum theory will preserve symmetries
provided a suitable measure factor can be found such that
\begin{equation}
\hat\delta([Df]\mu[f])=0.
\end{equation}
Moreover, the interaction vertices of the measure factor must be 
entire functions of the operator ${\cal K}$ and they must not destroy the 
FQFT finiteness.

In FQFT tree order, Green's functions remain local except for 
external lines which are unity on shell. It follows immediately that since 
on-shell tree amplitudes are unchanged by the regularization, $\hat W$ 
preserves all symmetries of $W$ on shell. Also all loops contain at least 
one regularizing propagator and therefore are ultraviolet finite. Shadow 
fields are eliminated at the classical level, for functionally integrating 
over them would produce divergences from shadow loops. Since shadow field 
propagators do not contain any poles there is no need to quantize the 
shadow fields. Feynman rules for ${\hat W}[f,h]$ are as simple as those for
local field theory.

\section{\bf Finite Quantum Yang-Mills and Kaluza-Klein Gravity Theory}

Let us now consider the finite quantization of the D-dimensional Yang-Mills
sector in D-dimensional Minkowski flat space. The gauge field strength 
$F_{aMN}$ is invariant under the familiar transformations:
\[
\delta A_{aM}=-\partial_M\theta_a+ ef_{abc}A_{bM}\theta_c.
\]

To regularize the Yang-Mills sector, we identify the kinetic operator
\[
{\cal K}_{ab}^{MN}=\delta_{ab}(\partial^2\eta^{MN}-\partial^M\partial^N).
\]
The regularized action is given by\cite{Kleppe2}
\begin{equation}
\hat{W}_{YM}[A]=\frac{1}{2}\int d^Dx\biggl\{\hat{A}_{aM}
{\cal K}^{MN}_{ab}\hat{A}_{bN}-B_{aM}[A]({\cal O}^{MN}_{ab})^{-1}
B_{bN}[A]\biggr\}
$$
$$
+W^I_{YM}[A+B[A]],
\end{equation}
where $B_{aM}$ is the Yang-Mills shadow field, which satisfies the 
expansion:
\[
B^M_a[A]={\cal O}^{MN}_{ab}\frac{\delta W^I_{YM}[A+B]}{\delta B^N_b}
\]
\[
={\cal O}^{MN}_{ab}ef_{bcd}[A_{Nc}\partial_KA^K_d+A_{cK}\partial_NA^K_d
-2A_{cK}\partial^KA^N_d]+O(e^2A^3).
\]
 
The regularized gauge symmetry transformation is
\[
\hat{\delta}_{\theta}A^M_a=({\cal E}^{MN}_{ab})^2
\biggl\{-\partial^M\theta_a+ef_{bcd}(A_{cN}+B_{cN}[A])\theta_d\biggr\}
\]
\[
-\partial^M\theta_a+({\cal E}^{MN}_{ab})^2ef_{bcd}(A_{cN}
+B_{cN}[A])\theta_d.
\]
The extended gauge transformation is neither linear nor local.

We functionally quantize the Yang-Mills sector using
\begin{equation}
\langle 0\vert T^*(O[A])\vert 0\rangle_{\cal E}=\int[DA]
\mu[A]({\rm gauge\, fixing})
O[{\hat A}]\exp(i\hat W_{\rm YM}[A]).
\end{equation}

To fix the gauge we use Becchi-Rouet-Stora-Tyutin (BRST)\cite{Becchi} 
invariance. The ghost structure of the BRST action comes from 
exponentiating the Faddeev-Popov determinant. Since the FQFT algebra fails 
to close off-shell, we need to introduce higher ghost terms into both the 
action and the BRST transformation. In Feynman gauge, the local BRST
Lagrangian is
\[
{\cal L}_{YM\,BRST}=-\frac{1}{2}\partial_MA_{aN}\partial^MA^N_a
-\partial^M\bar{\eta_a}\partial_M\eta_a+ef_{abc}\partial^M\bar{\eta}_a
A_{bM}\eta_c
\]
\[
+ef_{abc}\partial_MA_{aN}A^M_bA^N_c-\frac{1}{4}e^2f_{abc}f_{cde}
A_{aM}A_{bN}A^M_dA^N_c.
\]
It is invariant under the global symmetry transformation:
\[
\delta A_{aM}=(\partial_M\eta_a-ef_{abc}A_{bM}\eta_c)\delta\zeta,
\]
\[
\delta\eta_a=-\frac{1}{2}ef_{abc}\eta_b\eta_c\delta\zeta,
\]
\[
\delta\bar{\eta}_a=-\partial_MA^M_a\delta\zeta,
\]
where $\zeta$ is a constant anticommuting c-number. 

The gluon and ghost kinetic operators are
\[
{\cal K}^{MN}_{ab}=\delta_{ab}\eta^{MN}\partial^2,
\]
\[
{\cal K}_{ab}=\delta_{ab}\partial^2.
\]
The regularizing operators associated with the ghosts are
\[
\bar{\cal E}=\exp\biggl(\frac{\partial^2}{2\Lambda^2}\biggr),
\]
\[
\bar{\cal O}=\frac{\bar{\cal E}^2-1}{\partial^2}.
\]
The regularized BRST action is
\[
\hat{W}_{YM}[A,\bar{\eta},\eta]=\int d^Dz\biggl\{-\frac{1}{2}
\partial_N\hat{A}_{aM}\partial^N\hat{A}_a^M-\frac{1}{2}B_{aM}
\bar{\cal O}^{-1}B^M_a
\]
\[
-\partial^M\hat{\bar{\eta}}_a\partial_M\hat{\eta}_a
-\bar{\chi}_a\bar{\cal O}^{-1}\chi_a\biggr\}+W^I_{\rm YM}[A+B,\bar{\eta}
+\bar{\chi},\eta+\chi],
\]
where $\chi$ is the ghost shadow field.

The regularizing, nonlocal BRST symmetry transformation is
\[
\hat\delta A_{aM}=\bar{\cal E}^2\biggl\{(\partial_M\eta_a
+\partial_M\chi_a)-ef_{abc}(A_{bM}+B_{bM})(\eta_c+\chi_c)\biggr\}\delta\zeta,
\]
\[
\hat\delta\eta_a=-\frac{1}{2}ef_{abc}\bar{\cal E}^2(\eta_b+\chi_b)
(\eta_c+\chi_c)\delta\zeta,
\]
\[
\hat{\delta}\bar{\eta}_a=-\bar{\cal E}^2(\partial_MA^M_a
+\partial_MB^M_a)\delta\zeta.
\]
The full functional, gauge fixed quantization is now given by
\begin{equation}
\langle 0\vert T^*(O[A,\bar{\eta},\eta])\vert 0\rangle_{\cal E}=\int[DA]
[D\bar{\eta}][D\eta]
\mu[A,\bar{\eta},\eta]O[\hat{A},\hat{\bar{\eta}},\hat{\eta}]
$$
$$
\exp(i\hat{W}_{\rm YM}[A,\bar{\eta},\eta]). 
\end{equation}

Kleppe and Woodard\cite{Kleppe2} have obtained the invariant measure 
factor for the regularized Yang-Mills sector to first order in the 
coupling constant $e$:
\begin{equation}
\ln(\mu[A,\bar{\eta},\eta])=-\frac{1}{2}e^2f_{acd}f_{bcd}\int d^Dz
A_{aM}{\cal M}A^M_b+O(e^3),
\end{equation}
where
\[
{\cal M}=\frac{1}{2^D\pi^{D/2}}\int^1_0d\tau
\frac{\Lambda^{D-2}}{(\tau+1)^{D/2}}\exp\biggl(\frac{\tau}{\tau+1}
\frac{\partial^2}{\Lambda^2}\biggr)
\]
\[
\biggl\{\frac{2}{\tau+1}+1-D+2(D-1)\frac{\tau}{\tau+1}\biggr\}.
\]
The existence of a suitable invariant measure factor implies that the 
necessary Slavnov-Taylor identities also exist.

We shall now formulate in more detail the Kaluza-Klein gravitational sector
as a FQFT. This problem has been considered previously in the context of 
four-dimensional GR\protect\cite{Moffat,Moffat2,Moffat4}. We shall treat 
the theory as effectively being in D flat Minkowski dimensions, $D=4+d$.
A spacetime consisting of four flat Minkowski dimensions and $d$ 
circles of fixed radii $R=1/\mu_0$ cannot, in general, be equivalent to a 
flat D-dimensional spacetime. However, as the energy scale $\mu$ increases,
the effective length scale decreases, so that the fixed radius $R$ 
appears to become large, and the D-dimensional flat spacetime becomes a 
good approximation. In fact, FQFT can be formulated as a perturbative 
theory by expanding around any fixed, classical metric background, but 
for the sake of simplicity, we shall only consider in the 
following expansions about flat spacetime. As in ref.(\cite{Moffat4}), we 
will regularize the GR equations using the shadow field formalism.

We expand the local interpolating field ${\bf g}^{MN}=\sqrt{-g}
g^{MN}\quad (g={\rm Det}(g_{MN}))$ about Minkowski spacetime 
\begin{equation}
{\bf g}^{MN}=\eta^{MN}+\kappa\gamma^{MN}+O(\kappa^2).
\end{equation}
We separate the free and interacting 
parts of the action
\begin{equation}
W_{\rm grav}(g)=W^F_{\rm grav}(g)+W^I_{\rm grav}(g).
\end{equation}
The finite regularized gravitational action in FQFT is given by
\begin{equation}
\hat{W}_{\rm grav}(g,s)=W^F_{\rm grav}({\hat g})
-P_{\rm grav}(s)+W^I_{\rm grav}(g+s),
\end{equation}
where
\begin{equation}
{\hat g}={\cal E}^{-1}g,\quad P_{\rm grav}(s)
=\frac{1}{2}\int d^Dz{\cal F}(\sqrt{s},s_i{\cal O}_{ij}^{-1}s_j),
\end{equation}
$s$ denotes the graviton shadow field, and ${\cal F}$ denotes the detailed 
expansion of the metric tensor formed from the shadow field.

The quantum gravity perturbation theory is locally $SO(D-1,1)$ invariant 
(generalized, nonlinear field representation dependent transformations), 
unitary and finite to all orders in a way similar to the non-Abelian gauge
theories formulated using FQFT.  At the tree graph level all unphysical 
polarization states are decoupled and nonlocal effects will only occur in 
graviton and graviton-matter loop graphs. Because the gravitational tree 
graphs are purely local there is a well-defined classical GR limit.  The 
finite quantum gravity theory is well-defined in D real spacetime 
dimensions.

The graviton regularized propagator in a fixed de Donder gauge
is given in D-dimensional Minkowski space by\cite{Donder,Moff}
\[
D^{\rm grav}_{MNKL}
=(\eta_{MK}\eta_{NL}+\eta_{ML}\eta_{NK}
-\eta_{MN}\eta_{KL})
\]
\[
\biggl(\frac{-i}{(2\pi)^D}\biggr)\int d^Dk\frac{{\cal 
E}^2(k^2)}{k^2-i\epsilon} \exp[ik\cdot(z-z')],
\]
while the shadow propagator is
\[
D^{\rm shad}_{MNKL}
=(\eta_{MK}\eta_{NL}+\eta_{ML}\eta_{NK}
-\eta_{MN}\eta_{KL})
\]
\[
\biggl(\frac{-i}{(2\pi)^D}\biggr)\int d^Dk
\frac{[1-{\cal E}^2(k^2)]}{k^2-i\epsilon} \exp[ik\cdot(z-z')].
\]
In momentum space we have
\[
\frac{-i{\cal E}^2(k^2)}{k^2-i\epsilon}=-i\int^{\infty}_1\frac{d\tau}
{\Lambda^2_G}
\exp\biggl(-\tau\frac{k^2}{\Lambda^2_G}\biggr),
\]
and
\[
\frac{i({\cal E}^2(k^2)-1)}{k^2-i\epsilon}=-i\int_0^1\frac{d\tau}
{\Lambda^2_G}
\exp\biggl(-\tau\frac{k^2}{\Lambda^2_G}\biggr),
\]
where $\Lambda_G$ is the gravitational scale parameter.

We quantize by means of the path integral operation
\begin{equation}
\langle 0\vert T^*(O[g])\vert 0\rangle_{\cal E}=\int[Dg]
\mu[g]({\rm gauge\, fixing})
O[\hat g]\exp(i\hat W_{\rm grav}[g]).
\end{equation}
The quantization is carried out 
in the functional formalism by finding a measure factor
$\mu[g]$ to make $[Dg]$ invariant under the 
classical symmetry.
To ensure a correct gauge fixing scheme, we write $W_{\rm grav}[g]$
in the BRST invariant form with ghost fields; the ghost structure 
arises from exponentiating the Faddeev-Popov determinant. The algebra of 
gauge symmetries is not expected to close off-shell, so one needs to 
introduce higher ghost terms (beyond the normal ones) into both the 
action and the BRST transformation. The BRST action will be regularized 
directly to ensure that all the corrections to the measure factor are 
included.

\section{\bf Quantum Nonlocal Behavior in FQFT}

In FQFT, it can be
argued that the extended objects that replace point particles (the latter 
are obtained in the limit $\Lambda\rightarrow\infty$) cannot be probed 
because of a Heisenberg uncertainty type of argument.  The FQFT nonlocality
{\it only occurs at the quantum loop level}, so there is no noncausal 
classical behavior. In FQFT the strength of a signal propagated over an 
invariant interval $l^2$ outside the light cone would be suppressed by a 
factor $\exp(-l^2\Lambda^2)$.

Nonlocal field theories can possess non-perturbative instabilities. These
instabilities arise because of extra canonical degrees of freedom associated 
with higher time derivatives. If a Lagrangian contains up to $N$ time 
derivatives, then the associated Hamiltonian is linear in $N-1$ of the 
corresponding canonical variables and extra canonical degrees of freedom 
will be generated by the higher time derivatives. The nonlocal theory can 
be viewed as the limit $N\rightarrow\infty$ of an Nth derivative 
Lagrangian. Unless the dependence on the extra solutions is arbitrarily 
choppy in the limit, then the higher derivative limit will produce 
instabilities\protect\cite{Eliezer}. The condition for the smoothness of 
the extra solutions is that no invertible field redefinition exists which 
maps the nonlocal field equations into the local ones. String theory does 
satisfy this smoothness condition as can be seen by inspection of the 
S-matrix tree graphs. In FQFT the tree amplitudes agree with those of the 
local theory, so the smoothness condition is not obeyed.

It was proved by Kleppe and Woodard\protect\cite{Kleppe2} that the solutions 
of the nonlocal field equations in FQFT are in one-to-one correspondence 
with those of the original local theory. The relation for a generic field 
$v_i$ is \begin{equation}
v_i^{\rm nonlocal}={\cal E}^2_{ij}v^{\rm local}_j.
\end{equation}
Also the actions satisfy
\begin{equation}
W[v]={\hat W}[{\cal E}^2v].
\end{equation}
Thus, there are no extra classical solutions. 
The solutions of the regularized nonlocal Euler-Lagrange equations are 
in one-to-one correspondence with those of the local action. It follows 
{\it that the regularized nonlocal FQFT is free of higher derivative 
solutions, so FQFT can be a stable theory.}

Since only the quantum loop graphs in the nonlocal FQFT differ from 
the local field theory, then FQFT can be viewed as a non-canonical 
quantization of fields which obey the local equations of motion. Provided 
the functional quantization in FQFT is successful, then the theory does 
maintain perturbative unitarity.

\section{\bf A Resolution of The Higgs Hierarchy Problem and Quantum 
Gravity}

It is time to discuss the Higgs sector hierarchy problem\cite{Susskind}. 
The gauge hierarchy problem is related to the spin $0^+$ scalar field 
nature of the Higgs particle in the standard model with quadratic mass 
divergence and no protective extra symmetry at $m=0$. In standard point 
particle, local field theory the fermion masses are logarithmically 
divergent and there exists a chiral symmetry restoration at $m=0$.
 
Writing $m_H^2=m_{0H}^2+\delta m_H^2$, where
$m_{0H}$ is the bare Higgs mass and $\delta m_H$ is the Higgs 
self-energy renormalization constant, we get for the one loop Feynman 
graph in $D=4$ spacetime:
\begin{equation}
\delta m_H^2\sim \frac{g}{32\pi^2}\Lambda_C^2,
\end{equation}
where $\Lambda_C$ is a cutoff parameter. If we want to understand 
the nature of the Higgs mass we must require that

\begin{equation}
\delta m_H^2 \leq O(m_H^2),
\end{equation}
i.e. the quadratic divergence should be cut off at the mass scale of 
the order of the physical Higgs mass. Since $m_H\simeq \sqrt{2g}v$, 
where $v=<\phi>_0$ is the vacuum expectation value of the scalar field 
$\phi$ and $v=246$ GeV from the electroweak theory, then in order to keep 
perturbation theory valid, we must demand that $10\,{\rm GeV} 
\leq m_H \leq 350\,{\rm GeV}$ and we need \begin{equation}
\label{Higgscut}
\Lambda_C =\Lambda_{\rm Higgs}\leq 1\, {\rm TeV}, 
\end{equation}
where the lower bound on $m_H$ comes from the avoidance of washing out 
the spontaneous symmetry breaking of the vacuum.

Nothing in the standard model can tell us why (\ref{Higgscut}) should be true, 
so we must go beyond the standard model to solve the problem. $\Lambda_C$ 
is an arbitrary parameter in point particle field theory with no physical 
interpretation. Since all particles interact through gravity, then 
ultimately we should expect to include gravity in the standard model, so we
expect that $\Lambda_{\rm Planck}\sim 10^{19}$ GeV should be the natural 
cutoff. Then we have using (\ref{Higgscut}) and $g\sim 1$:
\[
\frac{\delta m_H^2(\Lambda_{\rm Higgs})}{\delta m_H^2
(\Lambda_{\rm Planck})} \approx \frac{\Lambda^2_{\rm Higgs}}
{\Lambda^2_{\rm Planck}}\approx 10^{-34}, \]
which represents an intolerable fine-tuning of parameters. This `naturalness' or 
hierarchy problem is one of the most serious defects of the standard model.

There have been two strategies proposed as ways out of the hierarchy
problem. The Higgs is taken to be composite at a scale $\Lambda_C\simeq 1$ TeV, 
thereby providing a natural cutoff in the quadratically divergent Higgs 
loops. One such scenario is the `technicolor' model, but it cannot be 
reconciled with the accurate standard model data, nor with the smallness of
fermion masses and the flavor-changing neutral current interactions. The other 
strategy is to postulate supersymmetry, so that the opposite signs of the 
boson and fermion lines cancel by means of the non-renormalization theorem.
However, supersymmetry is badly broken at lower energies, so we require 
that 
\[
\delta m_H^2\sim \frac{g}{32\pi^2}\vert\Lambda^2_{C\,{\rm bosons}}
-\Lambda^2_{C\,{\rm fermions}}\vert\leq 1\,{\rm TeV}^2,
\]
or, in effect
\[
\vert m_B-m_F\vert \leq 1\, {\rm TeV}.
\]
This physical requirement leads to the prediction that the supersymmetric 
partners of known particles should have a threshold $\leq1$ TeV. 

A third possible strategy is to introduce a FQFT formalism, and realize a field 
theory mechanism which will introduce a natural physical scale in the 
theory $\leq 1$ TeV, which will protect the Higgs mass from becoming large 
and unstable.

Let us consider the regularized scalar field FQFT Lagrangian in 
D-dimensional Minkowski space 
\begin{equation}
{\hat{\cal L}}_S=\frac{1}{2}\hat\phi(\partial^2-m^2)\hat\phi
-\frac{1}{2}\rho{\cal O}^{-1}\rho+\frac{1}{2}Z^{-1}\delta m^2(\phi+\rho)^2
-\frac{1}{24}g_0(\phi+\rho)^4,
\end{equation}
where $\phi=Z^{1/2}\phi_R$ is the bare field, $\phi_R$ is the renormalized 
field, $\hat\phi={\cal E}^{-1}\phi$, $\rho$ is
the shadow field, $m_0$ is the bare mass, $Z$ is the field strength 
renormalization constant, $\delta m^2$ is the mass renormalization constant
and $m$ is the physical mass. The regularizing operator is given by 
\begin{equation}
{\cal E}_m=\exp\biggl(\frac{\partial^2-m^2}{2\Lambda_H^2}\biggr),
\end{equation}
while the shadow kinetic operator is 
\begin{equation}
{\cal O}^{-1}=\frac{\partial^2-m^2}{{\cal E}_m^2-1}.
\end{equation}
Here, $\Lambda_H$ is the Higgs scalar field energy scale in FQFT.

The full propagator is
\begin{equation}
-i\Delta_R(p^2)=\frac{-i{\cal E}_m^2}{p^2+m^2-i\epsilon}
=-i\int_1^{\infty}\frac{d\tau}{\Lambda_H^2}\exp\biggl[-\tau
\biggl(\frac{p^2+m^2}{\Lambda_H^2}\biggr)\biggr],
\end{equation}
whereas the shadow propagator is
\begin{equation}
i\Delta_{\rm shadow}
=i\frac{{\cal E}_m^2-1}{p^2+m^2}=-i\int^1_0\frac{d\tau}{\Lambda_H^2}
\exp\biggl[-\tau\biggl(\frac{p^2+m^2}{\Lambda_H^2}\biggr)\biggr].
\end{equation}

Let us define the self-energy $\Sigma(p)$ as a Taylor series expansion 
around the mass shell
$p^2=-m^2$:
\begin{equation}
\Sigma(p^2)=\Sigma(-m^2)+(p^2+m^2)\frac{\partial\Sigma}{\partial p^2}(-m^2)
+{\tilde \Sigma}(p^2),
\end{equation}
where ${\tilde\Sigma}(p^2)$ is the usual finite part in the point 
particle limit $\Lambda_H\rightarrow\infty$.
We have
\begin{equation}
{\tilde\Sigma}(-m^2)=0,
\end{equation}
and 
\begin{equation}
\frac{\partial{\tilde\Sigma}(p^2)}{\partial p^2}(p^2=-m^2)=0.
\end{equation}

The full propagator is related to the self-energy $\Sigma(p^2)$ by
\begin{equation}
-i\Delta_R(p^2)=\frac{-i{\cal E}_m^2[1+{\cal O}\Sigma(p^2)]}{p^2+m^2
+\Sigma(p^2)}
=\frac{-iZ}{p^2+m^2+\Sigma_R(p^2)}.
\end{equation}
Here $\Sigma_R(p^2)$ is the renormalized self-energy which can be written 
as
\begin{equation}
\Sigma_R(p^2)=(p^2+m^2)\biggl[\frac{Z}{{\cal E}_m^2(1+{\cal O}\Sigma)}
-1\biggr]+\frac{Z\Sigma}{{\cal E}_m^2(1+{\cal O}\Sigma)}.
\end{equation}
The 1PI two-point function is given by
\begin{equation}
-i\Gamma_R^{(2)}(p^2)=i[\Delta_R(p^2)]^{-1}=\frac{i[p^2+m^2+\Sigma(p^2)]}
{{\cal E}_m^2[1+{\cal O}\Sigma(p^2)]}.
\end{equation}
Since ${\cal E}_m\rightarrow 1$ and ${\cal O}\rightarrow 0$ as $\Lambda_H
\rightarrow\infty$,
then in this limit
\begin{equation}
-i\Gamma_R^{(2)}(p^2)=i[p^2+m^2+\Sigma(p^2)],
\end{equation}
which is the standard point particle result.

The mass renormalization is determined by the propagator pole at $p^2=-m^2$ 
and we have
\begin{equation}
\Sigma_R(-m^2)=0.
\end{equation}
Also, we have the condition 
\begin{equation}
\frac{\partial\Sigma_R(p^2)}{\partial p^2}(p^2=-m^2)=0.
\end{equation}

The renormalized coupling constant is defined by the four-point function
$\Gamma^{(4)}_R(p_1,p_2,p_3,p_4)$ at the point $p_i=0$:
\begin{equation}
\Gamma_R^{(4)}(0,0,0,0)=g.
\end{equation}
The bare coupling constant $g_0$ is determined by
\begin{equation}
Z^2g_0=g+\delta g(g,m^2,\Lambda_H^2).
\end{equation}
Moreover,
\[
Z=1+\delta Z(g,m^2,\Lambda_H^2),
\]
\[
Zm_0^2=Zm^2-\delta m^2(g,m^2,\Lambda^2_H).
\]

A calculation of the scalar field mass renormalization gives\cite{Woodard2}:
\begin{equation}
\delta m^2=\frac{g}{2^{D+1}\pi^{D/2}}m^{D-2}
\Gamma\biggl(1-\frac{D}{2},\frac{m^2}{\Lambda_H^2}\biggr)+O(g^2),
\end{equation}
where $\Gamma(n,z)$ is the incomplete gamma function:
\begin{equation}
\Gamma(n,z)=\int_z^{\infty}\frac{dt}{t}t^n\exp(-t)=(n-1)\Gamma(n-1,z)
+z^{n-1}\exp(-z).
\end{equation}

We have
\begin{equation}
\Gamma(-1,z)=-E_i(z)+\frac{1}{z}\exp(-z),
\end{equation}
where $E_i(z)$ is the exponential integral
\[
E_i(z)\equiv \int^\infty_zdt\frac{\exp(-t)}{t}.
\]
For small $z$ we obtain the expansion
\begin{equation}
E_i(z)=-\ln(z)-\gamma+z-\frac{z^2}{2\cdot 2!}+\frac{z^3}{3\cdot 3!}-...,
\end{equation}
where $\gamma$ is Euler's constant. For large positive values of $z$, we 
have the asymptotic expansion
\begin{equation}
E_i(z)\sim\exp(-z)\biggl[\frac{1}{z}-\frac{1}{z^2}+\frac{2!}{z^3}-...\biggr].
\end{equation}
Thus, for small $m/\Lambda_H$ we obtain in $D=4$ spacetime:
\begin{equation}
\label{lambdast}
\delta m^2=\frac{g}{32\pi^2}
\biggl[\Lambda_H^2-m^2\ln\biggl(\frac{\Lambda_H^2}{m^2}\biggr)
-m^2(1-\gamma)+O\biggl(\frac{m^2}{\Lambda_H^2}\biggr)\biggr]+O(g^2),
\end{equation}
which is the standard quadratically divergent self-energy, obtained from 
a cutoff procedure or a dimensional regularization scheme.  

We have for $z\rightarrow\infty$:
\begin{equation}
\label{Gamma}
\Gamma(a,z)\sim 
z^{a-1}\exp(-z)\biggl[1+\frac{a-1}{z}
+O\biggl(\frac{1}{z^2}\biggr)\biggr] 
\end{equation}
so that for $m\gg\Lambda_H$, we get in D-dimensional space
\begin{equation}
\delta 
m^2\sim\frac{g}{2^{D+1}\pi^{D/2}}\biggl(\frac{\Lambda^D_H}{m^2}\biggr)
\exp\biggl(-\frac{m^2}{\Lambda_H^2}\biggr). \end{equation}
Thus, the Higgs self-energy one loop graph falls off exponentially fast for
$m\gg \Lambda_H$. 

The lowest order contributions to the graviton self-energy in FQFT will 
include the standard graviton loops, the shadow field graviton loops, the 
ghost field loop contributions with their shadow field counterparts, and 
the measure loop contributions. The calculated measures for regularized 
QED, first order Yang-Mills theory and all orders in $\phi^4$ and $\phi^6$ 
theories lead to self-energy contributions that are controlled by an 
incomplete $\Gamma$ function. For the regularized perturbative gravity 
theory the first order loop amplitude is 
\begin{equation}
\label{Aequation}
A_i=\Gamma\biggl(2-D/2,\frac{p^2}{\Lambda_G^2}\biggr)(p^2)^{D-2}
F_i(D/2).\quad (i=1,...,5)
\end{equation}
The dimensional regularization result is obtained by the replacement
\[
\Gamma\biggl(2-D/2,\frac{p^2}{\Lambda_G^2}\biggr)\rightarrow 
\Gamma(2-D/2),
\]
yielding the result\protect\cite{Capper,Veltman}
\begin{equation}
A_i\sim \Gamma(2-D/2)(p^2)^{D-2}F_i(D/2)\sim \frac{1}{\epsilon}
(p^2)^{D-2}F_i(D/2),
\end{equation}
where $\epsilon=2-D/2$ and $\Gamma(n)$ is the gamma function. 
Whereas the dimensional
regularization result is singular in the limit $\epsilon\rightarrow 
0$, the FQFT result is {\it finite in this limit} for a
fixed value of the parameter $\Lambda_G$, resulting in a finite 
graviton self-energy amplitude $A_i$. 

For D-dimensional spacetime, using (\ref{Gamma}) and (\ref{Aequation}), 
we obtain in the Euclidean momentum limit
\begin{equation}
A_i\sim \biggl(\frac{p^2}{\Lambda_G^2}\biggr)^{1-D/2}
\exp\biggl(-\frac{p^2}{\Lambda_G^2}\biggr)(p^2)^{D-2}F_i(D/2)\quad
(i=1,...,5).
\end{equation}
Thus, in the infinite Euclidean momentum limit the quantum graviton 
self-energy contribution damps out and quantum graviton corrections 
become negligible. It is often argued in the literature on quantum 
gravity that the gravitational quantum corrections scale as 
$\alpha_G =GE^2$, so that for sufficiently large values of the energy $E$, 
namely, of order the Planck energy, the gravitational quantum 
fluctuations become large. We see that in FQFT quantum gravity this may not
be the case, because for $\Lambda_G << M_{\rm Planck}\sim 10^{19}$ GeV, the
finite quantum loop corrections become negligible in the high energy limit.
Of course, the contributions of the tree graph exchanges of virtual 
gravitons can be large in the high energy limit, corresponding to strong 
classical gravitational fields. It follows that for high enough energies, 
a classical curved spacetime would be a good approximation.

In contrast to recent models of branes and strings in which the 
compactification scale is lowered to the TeV range\cite{Witten,Arkani}, we 
retain the classical GR gravitation picture and its Newtonian limit. It is 
perhaps a radical notion to entertain that quantum gravity becomes weaker 
as the energy scale increases towards the Planck scale $\sim 10^{19}$ Gev, 
but there is, of course, no known experimental reason why this should not 
be the case in nature. This would have important implications about the 
nature of singularities in gravitational collapse and the big bang 
scenario, for we could not appeal to quantum gravity to alleviate the 
singularity problem, but hope instead that a classical modification of GR 
occurs for very small distances $\leq 10^{-33}$ cm.

If we choose $\Lambda_H=\Lambda_G\geq$ 1 -5 TeV, then due to the damping of
the gravitational loop graphs and the scalar loop graphs in the Euclidean 
limit $p^2\gg\Lambda^2$, the Higgs sector is protected from large 
unstable radiative corrections and FQFT provides a solution to the Higgs 
hierarchy problem, without invoking low-energy supersymmetry or 
technicolor. The universal fixed FQFT scale $\Lambda_H$ corresponds to the 
fundamental length $\ell_H\leq 10^{-17}$ cm. For a Higgs mass much larger 
than 1 TeV, the Higgs sector becomes non-perturbative and we must be 
concerned about violations of unitarity\cite{Veltman2}.

\section{Kaluza-Klein Excitations Associated with Higher Dimensions}

In our D-dimensional gauge theory, we must concern ourselves with the 
restrictions imposed on the scale size, $\Lambda$, and the compactification
size, $R$, imposed by Kaluza-Klein excitations associated with the extra 
dimensions. Recently, Nath and Yamaguchi\cite{Nath}, and 
others\cite{Graesser}, have studied the constraints on the compactification
scale imposed by the Fermi constant, the fine structure constant, and the 
$W$ and $Z$ masses. Because of the importance of this issue for our theory,
we shall discuss these results here in some detail. Dienes et al. have 
caluculated the effects of infinite Kaluza-Klein towers on the "running" of
the gauge coupling constants\cite{Dienes2}. Let us now adapt their results 
to our FQFT formalism. 

As before, we shall treat our four-dimensional space as an approximately 
flat Minkowski spacetime, and evaluate the vacuum polarization diagram, 
including the effects of the Kaluza-Klein excitations on the loop graphs. 
Consider first the case of a single Dirac fermion. We can generalize this 
result to the case of a realistic chiral fermion model as a final step. In 
FQFT, the vacuum polarization tensor is the sum of three parts, 
$\Pi_1^{\mu\nu}(p), \Pi_2^{\mu\nu}(p)$, and $\Pi_3^{\mu\nu}(p)$ 
corresponding to the standard loop graph, a tadpole graph and a 
contribution from the invariant measure factor. We have
\begin{equation}
\Pi_1^{\mu\nu}=\Pi_1^T(p^2)\biggl(\eta^{\mu\nu}-\frac{p^\mu 
p^\nu}{p^2}\biggr)+\Pi^L_1(p^2)\frac{p^\mu p^\nu}{p^2},
\end{equation}
where $\Pi^T$ and $\Pi^L$ denote the transverse and longitudinal parts, 
respectively. The total transverse part is given by\cite{Moffat2}:
\begin{equation}
\label{vacuumstate}
\frac{\Pi^T(p^2)}{p^2}=-\frac{e^{(4)2}}{\pi^2}\sum^\infty_{n_i=-\infty}
\exp\biggl(-\frac{p^2}{\Lambda^2}\biggr)\int^{1/2}_0dx 
x(1-x)E_i\biggl[x(1-x)\frac{p^2}{\Lambda^2}+\frac{m_n^2}{\Lambda^2}\biggr],
\end{equation}
where $e^{(4)}$ is the four-dimensional coupling constant and
\[
\sum^\infty_{n_i=-\infty}\equiv 
\sum^\infty_{n_1=-\infty}\sum^\infty_{n_2=-\infty}...\sum^\infty_{n_d=-\infty},
\]
describes a summation over all Kaluza-Klein excitations with masses $m_n$, 
and $m_0$ is the energy of the ground state. Moreover, we see that 
$\Pi(p^2)$ vanishes exponentially fast in the Euclidean momentum region as 
$p^2\rightarrow\infty$.

The sum over the Kaluza-Klein states can be performed by using the Jacobi 
$\theta_3$ function\cite{Dienes2}:
\[
\theta_3(\tau)\equiv \sum^\infty_{n_i=-\infty}\exp(\pi i\tau n^2),
\]
where $\tau$ is a complex number. This function obeys the property
\[
\theta_3(-1/\tau)=\sqrt{-i\tau}\theta_3(\tau),
\]
where the branch of the square root with non-negative real part is chosen.
We obtain
\[
\frac{\Pi(p^2)}{p^2}=-\frac{e^{(4)2}}{\pi^2}\exp\biggl(-\frac{p^2}{\Lambda^2}\biggr)
\int^{1/2}_0dxx(1-x)E_i\biggl[x(1-x)\frac{p^2}{\Lambda^2}\biggr]
\biggl[\theta_3\biggl(\frac{it}{\pi R^2}\biggr)\biggr]^d.
\]
We then get
\begin{equation}
\biggl[\frac{\Pi(p^2)}{p^2}\biggr]_{p^2=0}=-\frac{e^{(4)2}}{12\pi^2}
\int^{r^2/\mu_0^2}_{r^2/\Lambda^2}\frac{dt}{t}\biggl[\theta_3\biggl(\frac{it}{\pi
R^2} \biggr)\biggr]^d,
\end{equation}
where 
\[
r^2=\frac{\pi}{(X_d)^{2/d}}
\]
and
$$
X_d=\frac{\pi^{d/2}}{\Gamma(1+d/2)}
$$
relates our FQFT scale $\Lambda$ to the underlying physical mass scales. It
is to be noted that in contrast to the standard cut-off technique, the FQFT
calculation of the vacuum polarization is fully gauge invariant and unitary
and $\Lambda$ represents a {\it physical} scale in the theory.

For our full chiral gauge theory assuming that the Kaluza-Klein excitations
arise from the gauge bosons and Higgs fields only, we obtain
\begin{equation}
\biggl[\frac{\Pi(p^2)}{p^2}\biggr]_{p^2=0}
=\frac{e_i^{(4)2}(b_i-\tilde{b}_i)}{8\pi^2}\ln\biggl(\frac{\Lambda}{\mu_0}\biggr)
$$
$$
+\frac{e_i^{(4)2}\tilde{b}_i}{16\pi^2}
\int^{r^2/\mu_0^2}_{r^2/\Lambda^2}\frac{dt}{t}
\biggl[\theta_3\biggl(\frac{it}{\pi R^2}\biggr)\biggr]^d,
\end{equation}
where the $b_i$ are the one-loop beta-functions for the zero modes, while 
the $\tilde{b}_i$ denote the beta-functions associated with the 
Kaluza-Klein excitations.

We can now obtain the scaling behaviour of our gauge coupling constants
\begin{equation}
\alpha_i^{-1}(\Lambda)=\alpha^{-1}_i(\mu_0)-\biggl(\frac{b_i-\tilde{b}_i}{2\pi}\biggr)
\ln\biggl(\frac{\Lambda}{\mu_0}\biggr)
$$
$$
-\frac{\tilde{b}_i}{4\pi}\int^{r^2/\mu_0^2}_{r^2/\Lambda^2}
\frac{dt}{t}\biggl[\theta_3\biggl(\frac{it}{\pi R^2}\biggr)\biggr]^d,
\end{equation}
where $\alpha_i\equiv e_i^{(4)2}/4\pi$. This gives the
running of the gauge coupling constants as obtained by Dienes et 
al.\cite{Dienes2}. The important result emerges that the Kaluza-Klein 
exitations convert the standard logarithmic scaling of the gauge coupling 
constants to a power law running behaviour.

By imposing matching conditions that the uncorrected value of the effective
four-dimensional coupling constant $\alpha_i$ must agree with the value of 
the four-dimensional coupling $\alpha_i(\mu_0)$ at the scale $\mu_0$, we 
get
\begin{equation}
\alpha_i^{-1}(\Lambda)=\alpha^{-1}_i(M_Z)-\frac{b_i}{2\pi}
\ln\biggl(\frac{\Lambda}{M_Z}\biggr)
+\frac{\tilde{b}_i}{2\pi}\ln\biggl(\frac{\Lambda}{\mu_0}\biggr)
-\frac{\tilde{b}_iX_d}
{2\pi d}\biggl[\biggl(\frac{\Lambda}{\mu_0}\biggr)^d-1\biggr], 
\end{equation}
where $M_Z$ is the mass of the Z-boson. This result is valid for all 
$\Lambda\geq \mu_0$. The Kaluza-Klein excitations cause {\it an 
acceleration of the unification of the gauge coupling constants} 
$\alpha_i$; this leads to the possibility of having gauge coupling 
unification at energies well below the usual GUT scale of $10^{16-19}$ GeV.

Let us consider the unregulated
interaction Lagrangian, describing the coupling of fermions to zero modes 
and to the Kaluza-Klein modes: 
\begin{equation} {\cal L}_{\rm int} 
=e^{(4)}_iJ^\mu(A_{\mu i}+\sqrt{2}\sum^{\infty}_{n=1}A^{(n)}_{\mu i}), 
\end{equation} 
where $A_{\mu i}$ are the zero modes and $A^{(n)}_{\mu i}$ are the 
Kaluza-Klein modes, respectively. Integrating out the $W$ boson and its 
Kaluza-Klein excitations gives for the effective standard model Fermi 
constant\cite{Nath,Graesser} 
\begin{equation} G^{\rm eff}_F=G^{\rm 
SM}_FK_d\biggl(\frac{M_W^2}{M_R^2}\biggr), 
\end{equation} where $d=D-4$ and
$K_d$ is \[
K_d(s)=\int^\infty_0 dt\exp(-t)\biggl[\theta_3\biggl(\frac{it}{s\pi}\biggr)
\biggr]^d.
\]
This integral diverges for $d > 1$, but a convergent result is obtained 
in regularized FQFT, corresponding to a truncation of the Kaluza-Klein states when the 
masses exceed the FQFT scale $\Lambda$, associated with the regulated 
Lagrangian.

Nath and Yamaguchi obtain the ratio of the Kaluza-Klein contribution to the
Fermi constant to the standard model value of the Fermi constant for 
$d\geq 3$: 
\begin{equation}
\frac{\Delta G_F^{KK}}{G_F^{\rm SM}}\simeq \biggl(\frac{d}{d-2}\biggr)
\frac{\pi^{d/2}}{\Gamma(1+d/2)}
\biggl(\frac{\Lambda}{M_R}\biggr)^{d-2}\biggl(\frac{M_W}{M_R}\biggr)^2.
\end{equation}

The gauge coupling evolution constrains $M_R$ and $\Lambda$ for TeV scale 
unification. Let us adopt the evolution equation\cite{Dienes2} used 
by Nath and Yamaguchi:
\begin{equation} 
\alpha_i(M_Z)=\frac{1}{\alpha_U}+\frac{b_i}{2\pi}\ln\biggl( 
\frac{M_R}{M_Z}\biggr)-\frac{b_i^{KK}}{2\pi}\ln\biggl(\frac{\Lambda}{M_R}
\biggr)
+\Delta_i,
\end{equation}
where $\alpha_U$ is the effective GUT coupling constant, $b_i=(-3,1,33/5)$ 
for $SU(3)_c\times SU(2)\times U(1)$, $b^{KK}_i=(-6,-3,3/5)$ are the 
$b_i$ minus the fermion sector contribution which has no Kaluza-Klein 
excitations, and $\Delta_i$ are the Kaluza-Klein corrections. Given $d$ and
$M_R$ the unification of $\alpha_1$ and $\alpha_2$ fixes $\Lambda/M_R$. For
the cases of $d$ equal to 2, 3 and 4 extra dimensions, the Nath-Yamaguchi 
analysis produces the lower limits on $M_R$ of 3.5 TeV, 5.7 TeV and 7.8 
TeV. Thus, the observation of Kaluza-Klein excitations may be possible at 
the Large Hadron Collider (LHC). 

Nath and Yamaguchi\cite{Nath} have also considered the constraints arising 
from an analysis of $g_\mu-2$ for extra $d$ dimensions. These constraints 
on $M_R$ at the $2\sigma$ level are $M_R > 1.6$ TeV for $d=1$, $M_R > 3.5$ 
TeV for $d=2$, $M_R > 5.7$ TeV for $d=3$ and $M_R > 7.8$ TeV for $d=4$.

\section{Proton Decay Lifetime and Unified Theory Phenomenology}

The problem of proton decay must be considered in the context of the 
unified models.  When both colour triplet quarks and colour singlet leptons are 
assigned to the same irreducible representation of a symmetry group, then 
there exist vector bosons (leptoquarks) that transform leptons into quarks.
Unless there exists a conserved quantum number $A$ for which the proton is 
the lowest mass state with $A=1$, then the proton can decay. We must 
guarantee within our unified field theory that the proton is stabilized 
sufficiently to not disagree with the experimental bounds on its lifetime,
$\tau_p\geq 10^{32}$ yrs. For example, as is well-known, the conventional 
$SU(5)$ model of Georgi and Glashow\cite{Glashow} has been eliminated by 
the experimental bound on the decay rate. The problem becomes much more 
severe when we contemplate a compactification scale $M_c$ of order 1-10 
TeV.

In our theory, there exist several energy scales to consider. There is the 
Yang-Mills scale $\Lambda_{\rm YM}$, associated with the Yang-Mills 
Lagrangian, ${\cal L}_{\rm YM}$, the Higgs scalar field scale $\Lambda_H$, 
and the gravitational scale $\Lambda_G$. In addition, there is the 
aforementioned compactification scale $M_c$. Let us first consider the 
possibility that the Yang-Mills and the compactification scales are large, 
$\Lambda_{\rm YM}\sim M_c\geq 10^{16}$ GeV.

For the $SO(10)$ model, we choose the left-right symmetric 
$SU(2)$ Pati-Salam breaking pattern \cite{Pati}:
\[
SO(10)\rightarrow SU(4)\times SU(2)_L\times SU(2)_R\rightarrow SU(3)\times 
SU(2)_L\times SU(2)_R\times U'(1)_{B-L}
\]
\[
\rightarrow SU(3)\times SU(2)\times 
U(1)\rightarrow SU(3)\times U(1)_{\rm em}.
\]
Here, the sequence of characteristic Higgs potentials are described by the 
representations $\underline{54}, \underline{45}, \underline{16}$ and 
$\underline{10}$. At one loop level, the renormalization group equations 
are for $\mu < M_c \sim\Lambda_{\rm YM}$\cite{Dienes2,Ross}
\begin{equation}
\sin^2\theta_W(M_W)=\frac{3}{8}-\frac{11}{3}\frac{\alpha(M_W)}{\pi}
\biggl[\frac{5}{8}\ln\biggl(\frac{M_U}{M_W}\biggr)-\frac{3}{8}
\ln\biggl(\frac{M_U}{M_S}\biggr)\biggr],
\end{equation}
and
\begin{equation}
\frac{\alpha_{\rm em}(M_W)}{\alpha_3(M_W)}=\frac{3}{8}-\frac{11}{8}
\frac{\alpha(M_W)}{\pi}\biggl[3\ln\biggl(\frac{M_U}{M_W}\biggr)
-\ln\biggl(\frac{M_U}{M_S}\biggr)\biggr],
\end{equation}
where $M_U$ denotes the mass of the exchanged boson that breaks $SO(10)$,
while $M_W$ is the weak scale mass. We have assumed that $M_C\sim M_U$,
where $M_C$ is the mass of the exchanged boson that breaks $SU(4)\times 
SU(2)_L\times SU(2)_R$ to $SU(3)\times SU(2)_L\times SU(2)_R\times 
U'(1)_{B-L}$. Moreover, $M_S$ denotes the mass associated with the breaking
of $SU(3)\times SU(2)_L\times SU(2)_R\times U'(1)_{B-L}$ to $SU(3)\times
SU(2)\times U(1)$. The term proportional to $\ln\biggl(M_U/M_W\biggr)$ 
corresponds to the low energy group $SU(3)\times SU(2)\times U(1)$. We can 
choose $M_S$ to keep the prediction for $\alpha(M_W)/\alpha_3(M_W)$ fixed 
at the $SU(5)$ value, while varying $M_U$ from the usual $SU(5)$ value for 
$M_X$. This increases $\sin^2\theta_W(M_W)$ by approximately $0.005
\ln\biggl(M_U/M_X\biggr)$. We can still obtain reasonable values for 
$\sin^2\theta_W$ for a large variation of $M_U$. Consequently, the proton 
decay lifetime, which satisfies
\begin{equation}
\tau_p\propto 10^{30}\,{\rm yrs}\biggl(\frac{M_U}{5\times 10^{14}\,{\rm 
GeV}}\biggr)^4
\end{equation}
can be made to satisfy the experimental bounds. 

This illustrates the freedom in building an $SO(10)$ model, because of the 
uncertainties in the values of $M_U$ and $\tau_p$, once we go beyond the 
simplest, minimal symmetry breaking scheme. However, these models do 
possess the possibility of having very light scales of intermediate 
unification, which may be associated with baryon and lepton number 
violating processes in other than proton channel decays.

We must now turn our attention to the Higgs hierarchy problem. With the 
foregoing scenario, we can choose for the Higgs scale, $\Lambda_H\sim 1$ 
TeV, which guarantees in FQFT (see Sect.7) that the {\it radiative loop 
contributions to the scalar Higgs self-energy are controlled}, leaving the 
tree graphs untouched at higher energies, which can mediate Higgs 
spontaneous symmetry breaking. We have the possibility of reducing the 
graviton quantum loop contributions to the TeV scale by choosing 
$\Lambda_G\sim 1$ Tev, or choosing a much higher value for $\Lambda_G$, 
e.g. the Planck scale $\sim 10^{19}$ GeV.

This scenario does solve the Higgs hierarchy problem, while simultaneously
ensuring a sufficiently stable proton, but the extreme difference between 
the Yang-Mills scale $\Lambda_{\rm YM}$ and the Higgs scalar field scale 
$\Lambda_H$ does raise questions about "naturalness". However, we stress
that these energy scales are associated with {\it physical} scales in FQFT,
and not with arbitrary cut-offs.

Let us consider next a scenario in which $\Lambda_{\rm YM}\sim M_c\sim 
\Lambda_H\sim \Lambda_G\sim$ 1-10 TeV. This produces, at first sight, an 
attractive physical picture in which we can contemplate observing new 
physics with the next generation of accelerators. The classical GR theory 
is left unchanged, because only the graviton quantum loop graphs are 
reduced to a scale of 1-10 TeV, leaving classical GR intact to all 
energies. The gauge coupling constant will satisfy a power law behaviour 
when $\Lambda_{\rm YM}\sim M_c > \mu$, so gauge coupling unification is 
accelerated, as explained in Sect 8. There is no Higgs hierarchy problem in
four dimensions, and for our $SO(18)$ unification scheme {\it we obtain the
correct prediction of three chiral families and the standard model in four 
dimensions}. However, we are faced with the {\it b${\hat e}$te noire} of 
potentially fast proton decay.

There are two possible scenarios that we can adopt to circumvent the 
problem of baryon and lepton number violation. The first is to introduce a 
global $U(1)$ symmetry. The conservation of $A$ does not carry with it a 
long-range force, so even though $A$ is an additive quantum number like 
charge $Q$, it cannot be a generator of a local $U(1)$. However, if a 
global $U(1)$ and a local $U(1)$ are both broken in such a way that a 
linear combination of the generators is conserved, then the vector boson 
acquires a mass and the unbroken linear combination yields an exact 
conservation law\cite{Gell-Mann2}. Let $X$ be the generator of the local 
$U(1)$ and $Z$ the generator of the global $U(1)$. We choose $Z=0$ for the 
fermions and assign all fermions to a single irreducible representation (as
is the case for our $SO(10)$ in four dimensions). The Higgs fields have 
non-zero values of $Z$, and if $Z$ and $X$ are broken and their sum is 
conserved, then some of the scalar fields will acquire non-zero $A$, 
resulting in a heavy boson and $A=X$ in the fermion sector. Thus, our 
$SO(10)$ theory has a stable proton. It is possible to break the $Z$ and 
$X$ so that no ``weird" fermions exist, but in general there will exist 
"weird" fermions at higher energies.  

Of course, the introduction of a global $U(1)$ symmetry into the theory is 
considered by some to be an unpalatable way of solving the proton decay 
problem. It goes against the spirit of local gauge symmetries. It means 
that if we rotate a proton in our living room, then exactly the same 
rotation is required to be performed in the Andromeda galaxy. There is also
the danger that gravitational interactions will induce fast proton decay in
higher-dimensional operators. Moreover, there is the more abstract issue 
that black holes violate all non-gauged symmetries.

Another possible scenario to evade the proton decay issue has recently been
proposed by Arkani-Hamed and Schmaltz\cite{Schmaltz}. They ``stick" the 
standard model fermions at different points on domain walls in the 
extra $d$ dimensions. The couplings between them are suppressed due to the 
exponentially small overlaps of their wave functions. We can adopt this 
mechanism in our higher-dimensional field theory. The model can be simply 
visualized in one extra dimension, in which the gauge fields and the Higgs 
fields are allowed to propagate inside the wall, while the fermions are 
constrained to different points in the wall. The fermion wave functions are
described by narrow Gaussian functions. The long-distance four-dimensional 
theory can have exponentially small Yukawa couplings, generated by the 
small overlap between left- and right-handed fermion wave functions. This 
can lead to an exponentially suppressed proton decay rate, if the quarks 
and leptons are localized to separate ends of the wall. This avoids the
issue of inventing symmetries in the theory which protect the proton from a
fast decay rate. Since this mechanism relies only on the dynamics of the 
wall geometry and the placements of the fermions on the walls in the extra 
dimensions, one may develop an uneasy feeling that the mechanism is 
somewhat contrived, but at present there appears to be no obvious critical 
reason why this could not be an acceptable way to resolve the proton decay 
problem.

\section{\bf Conclusions}

A higher-dimensional unified field theory based on a 
Kaluza-Klein-Yang-Mills-Higgs action and a ground state $M^4\times B$ is 
developed. The gauge group has a topologically non-trivial 
sub-group associated with the compact internal space, i.e. a 
Dirac monopole is inserted into the latter group to guarantee that the 
compactification to four dimensions retains the chiral non-vector-like 
property of the fermions. We choose the minimal, anomaly free model 
$SO(18)$ in twelve dimensions with the breaking to $SO(8)\times SO(10)$, 
leading to a four-dimensional group $SO(10)$ GUT model, which contains the 
standard model $SU(3)_c\times SU(2)\times U(1)$ and predicts {\it three 
families of quarks and leptons}. The chiral nature of the fermion 
representations is guaranteed in the dimensional reduction process by 
making $SO(8)$ topologically non-trivial with the reduction 
$SO(8)\rightarrow Sp(4)\times SU(2)$. The fermions have a non-vanishing 
chirality number, whereby the left-handed quarks and leptons have the 
correct physical interpretation.

The problem of the stability of Kaluza-Klein theories can be solved in 
our model by means of the supplementary Yang-Mills fields or the magnetic 
monopole that exists in the internal space, which can generate the 
repulsive forces necessary to balance the gravitational forces.

The important gauge hierarchy problem, associated with the Higgs sector, is
solved by the exponential damping of the Higgs self-energy in the Euclidean
$p^2$ domain for $p^2 > \Lambda_H^2$, and for a $\Lambda_H$ scale in the 
TeV range.  

The unified field theory will have three coupling constants, 
namely, the gravitational constant $\bar G$, the gauge coupling 
constant $e$, and the scalar Higgs coupling constant g, which have 
to be rescaled in four dimensions. Dimensional reduction to four dimensions
can explain charge quantization in terms of the compactification scale $R$.
In string theory, the coupling constants are determined by the string 
dilaton scalar field, so in principle there are no arbitrary coupling 
constants. However, this presupposes that one knows the {\it complete} 
solution to the dynamics of the string equations. It is difficult to see 
how the determination of the coupling constants in string theory can be 
implemented in practice, particularly, since it would appear that only 
non-perturbative solutions can be obtained in M-theory.

The critical issue of the finiteness of quantum gravity perturbation theory
in D dimensions is solved by applying the FQFT formalism. The nonlocal 
quantum loop interactions reflect the non-point-like nature of the 
field theory, although we do not specify the nature of the extended object 
that describes a particle. Thus, as with string theories, the point-like 
nature of particles is ``fuzzy" in FQFT for energies greater than the scale
$\Lambda$. One of the features of superstrings is that they provide a 
mathematically consistent theory of quantum gravity, which is ultraviolet 
finite and unitary. FQFT focuses on the basic mechanism behind string 
theory's finite ultraviolet behavior by invoking a suppression of bad 
vertex behavior at high energies, without compromising perturbative 
unitarity and gauge invariance. FQFT provides a mathematically consistent 
theory of quantum gravity at the perturbative level. If we choose 
$\Lambda_G\sim$ 1-10 TeV, then quantum radiative corrections to the 
classical tree graph gravity theory are perturbatively negligible to all 
energies greater than $\Lambda_G$ including the Planck energy. If, on the 
other hand, we choose $\Lambda_G\sim M_{\rm Planck}$, then we are forced to
seek a non-perturbative FQFT quantum gravity formalism at the Planck scale.

Our solution of the finiteness of quantum gravity is based on a {\it gauge 
field theory}, which allows us to go off the mass shell when calculating 
vertex operators. It is generally accepted that there is no self-consistent
string field theory, because such a theory would correspond to a $\phi^3$ 
theory with a Hamiltonian unbounded from below, causing the field theory to
be unstable. Our FQFT can be a stable theory as was shown by Kleppe and 
Woodard\cite{Kleppe2}. However, our higher-dimensional field theory may be
linked to a final, realized stable version of M-theory, for the finiteness 
of FQFT owes its existence to an ultraviolet suppression mechanism 
akin to that of string theory, and a field theory version of M-theory could
possess a structure similar to our FQFT. Superstring theory and its
offspring appear to lead to fundamental changes in our understanding of 
space and time\cite{Hooft}. In particular, our everyday notions of 
causality may be altered at high energies and small distances. Our 
introduction of nonlocal interaction Lagrangians in FQFT may be the 
modification of local, point particle field theory at the quantum level 
that is needed to achieve a mathematically consistent theory of quantum 
gravity.

Supersymmetry is required if we wish to unify the particle spins of bosons 
and fermions. This would be a mathematically beautiful achievement. We 
could incorporate supersymmetry in our higher-dimensional FQFT, and indeed 
in contrast to, for example, dimensional regularization, FQFT respects 
continuous supersymmetry gauge transformations to all orders of 
perturbation theory. However, supersymmetry partners have not, as yet, been
observed and as we have demonstrated, FQFT can resolve the Higgs hierarchy 
problem without supersymmetry, thereby removing the primary reason for 
promoting supersymmetry at the phenomenological level.

Because we are able to lower the compactification scale $M_c$ to the TeV 
energy range, we anticipate that Kaluza-Klein excitations will be 
observable at these energies by the LHC. In order to distinguish these 
Kaluza-Klein signatures from the signatures of supersymmetry partners or 
other exotic physics, we must analyze further the decay properties of the 
Kaluza-Klein modes, so as to select possible unique features associated 
with the excitations.

We have adopted the idea of reducing quantum gravity to lower energies by 
choosing the quantum gravity scale $\Lambda_G$ in the TeV range, or even at
much lower energies. This is in accord with the recent interesting idea 
that the gravitational scale could be in the TeV range\cite{Arkani}. 
However, in contrast to the work in ref. (8), we only lower the energy 
scale of the quantum gravity loops through the choice of the scale 
$\Lambda_G$, without affecting the classical tree graphs which sum to give 
local and causal, classical Newtonian and GR theories. 

We have not considered the
implications of our theory for cosmology and black holes, nor have we 
concerned ourselves with the important problem of the cosmological 
constant. These issues will be addressed in future work. 
\vskip 0.2 true in
{\bf Acknowledgments} \vskip 0.2 true in This work was supported by the 
Natural Sciences and Engineering Research Council of Canada.
\vskip 0.5 true in


\begin{thebibliography}{100}

\bibitem{Altarelli} G. Altarelli, hep-ph/9809532.

\bibitem{Schwarz} M. Green and J. H. Schwarz, Phys. Lett. B{\bf 149}, 
117 (1984); D. Gross, J. Harvey, E. Martinec and R. Rohm, Phys. Rev. Lett. 
{\bf 54}, 502 (1985); P. Candelas, G. Horowitz, A. Strominger and E. 
Witten, Nucl. Phys. B{\bf 258}, 46 (1985).; M. Green, J. H. Schwarz and E. 
Witten, {\it Superstring Theory}, vols. I and II, Cambridge University 
Press (1986); J. Polchinski, {\it String Theory}, vols. I and II, Cambridge
University Press, 1998; L. E. Ibanez, hep-ph/9901292.

\bibitem{Duff} For reviews see:``Lectures on Superstring and M-theory 
Dualities", hep-th/9607201 (1996); J. Polchinski, TASI Lectures on 
D-branes," hep-th/9611050 (1996); P. Townsend, ``Four Lectures on 
M-theory," hep-th/9612121 (1996); A. Sen, ``Unification of String 
Dualities", hep-th/9609176 (1996); M. Duff, hep-th/9805177.

\bibitem{Witten} E. Witten, Nucl. Phys. B{\bf 471}, 135 (1996); J. D. 
Lykken, Phys. Rev. D{\bf 54}, 3693 (1996).

\bibitem{Dienes} K. Dienes, E. Dudas and T. Gherghetta, Phys. Lett. B{\bf 
436}, 55 (1998); D. Ghilencea and G. G. Ross, hep-ph/9809217
 
\bibitem{Shiu} G. Shiu and S. H. Henry Tye, Phys. Rev. D{\bf 58}, 106007 
(1998); Z. Kakushadze, hep-th/981193.

\bibitem{Dienes2} K. Dienes, E. Dudas, and T. Gherghetta, Nucl. Phys.
B{\bf 537}, 47 (1999).
 
\bibitem{Arkani} I. Antoniadis, Phys. Lett. B{\bf 246}, 377 (1990); I. 
Antoniadis, N. Arkani-Hamed, S. Dimopolous, and G. Dvali, Phys. Lett. 
B{\bf 429}, 263 (1998); N. Arkani-Hamed, S. Dimopoulos, and G. Dvali, 
Phys. Rev. D{\bf 59}, 105002 (1999).

\bibitem{Moff} J. W. Moffat, hep-ph/9802228.

\bibitem{Moffat} J. W. Moffat, Phys. Rev. D{\bf 41}, 1177 
(1990).

\bibitem{Moffat2} D. Evens, J. W. Moffat, G. Kleppe and R. P. Woodard, 
Phys. Rev. D{\bf43}, 49 (1991).

\bibitem{Moffat3} J. W. Moffat and S. M. Robbins, Mod. Phys. Lett. 
A{\bf 6}, 1581 (1991).

\bibitem{Woodard} G. Kleppe and R. P. Woodard, Phys. Lett. B{\bf 253}, 
331 (1991).

\bibitem{Kleppe2} G. Kleppe and R. P. Woodard, Nucl. Phys. B{\bf 388}, 
81 (1992).

\bibitem{Hand} B. Hand, Phys. Lett. B{\bf 275}, 419 (1992).

\bibitem{Woodard2} G. Kleppe and R. P. Woodard, Ann. of  Phys. 
{\bf 221}, 106 (1993).

\bibitem{Clayton} M. A. Clayton, L. Demopolous and J. W. Moffat, 
Int. J. Mod. Phys. A{\bf 9}, 4549 (1994).

\bibitem{Moffat4} J. W. Moffat, hep-th/9808091. Talk given at
the XI International Conference on Problems in Quantum Field
Theory, Dubna, Russia, July 13-17, 1998. To be published in the Proceedings
by World Scientific, Singapore.
 
\bibitem{Kaluza} Th. Kaluza, Sitzungsber. 
Preuss. Akad. Wiss. Berlin, Math. Phys. Kl 966 (1921); O. Klein, Z. Phys. 
{\bf 37}, 895 (1926), Ark. Mat. Astron. Fys. B {\bf 34}A (1946); A. 
Einstein and P. Bergman, Ann. Math. {\bf 39}, 683 (1938); J. Rayski, {\bf 
27}, 89 (1965); R. Kerner, Ann. Inst. Henri Poincar\'e, {\bf 9}, 143 
(1968); E. Witten, Nucl. Phys. B{\bf 186}, 412 (1981). For reprints of 
papers on Kaluza-Klein theories and a more complete list of references, 
see: {\it Modern Kaluza-Klein Theories}, eds. T. Appelquist, A. Chodos, and
P. G. O. Freund, Frontiers in Physics, Addison-Wesley Pub. Co., California,
1987; L. O'Raifeartaigh and N. Straumann, hep-ph/9810524, to be published 
in Rev. Mod. Phys.

\bibitem{Wetterich} C. Wetterich, Nucl. Phys. B{\bf 242}, 473 (1984). 

\bibitem{Hirzebruch} M. F. Atiyah and F. Hirzebruch, in {\it Essays on 
Topology and Related Topics}, edited by A. Haefliger and R. Narasimhan. 
Springer, Berlin, 1970, p.18.  

\bibitem{Cremmer} E. Cremmer and J. Scherk, Nucl. Phys. B{\bf 108}, 409 
(1977); Z. Horvath, L. Palla, E. Cremmer, and J. Scherk, Nucl. Phys. B{\bf 
127}, 57 (1977).

\bibitem{Luciani} J. F. Luciani, Nucl. Phys. B{\bf 135}, 111 (1978).

\bibitem{Glashow} H. Georgi and S. L. Glashow, Phys. Rev. Lett. {\bf 32}, 
438 (1974).
 
\bibitem{Fritzsch} H. Fritzsch and P. Minkowski, Ann. of Phys. {\bf 93}, 
193 (1975); H. Georgi, in Particles and Fields--1974, Proceedings of the 
1974 Meeting of the APS Division of Particles and Fields, edited by C. 
Carlson (AIP, NEW York, 1975).

\bibitem{Witten2} E. Witten, Princeton University preprint No. PHY80-19754,
reprinted in {\it Modern Kaluza-Klein Theories}, eds. T. Appelquist, A. 
Chodos, and P. G. O. Freund, Frontiers in Physics, Addison-Wesley Pub. Co.,
California, 1987.
 
\bibitem{Lichnerowicz} A. Lichnerowicz, C. R. Acad. Sci. Paris, Serie A-B
{\bf 257}, 7 (1963).

\bibitem{Slansky} G. Chapline and R. Slansky, Nucl. Phys. B{\bf 209}, 461 
(1982).

\bibitem{Salam} S. Randjbar-Daemi, A. Salam and J. Strathdee, Nucl. Phys. 
B{\bf 214}, 491 (1983); S. Randjbar-Daemi and R. Percacci, Phys. Lett. 
B{\bf 117}, 41 (1982).
 
\bibitem{Gell-Mann} M. Gell-Mann, P. Ramond, and R. Slansky, 
{\it Supergravity}, edited by P. van Nieuwenhuizen and D. Z. Friedman, 
North-Holland Publishing Company, 1979.
 
\bibitem{Zee} F. Wilczek and A. Zee, Phys. Rev. D{\bf 25}, 553 (1982).

\bibitem{Becchi} C. Becchi, A. Rouet, and R. Stora, Comm. Math. Phys.
{\bf 42}, 127 (1975); I. V. Tyutin, Lebedev Institute preprint N39 (1975).
 
\bibitem{Feynman} The issue of uniqueness in our interpretation 
of nature has always been a source of debate. Richard Feynman, shortly 
before his death, made the following comment: "My feeling has been--and I 
could be wrong--that there is more than one way to skin a cat. I don't 
think that there's only one way to get rid of infinities. The fact that a 
theory gets rid of infinities is to me not a sufficient reason to 
believe its uniqueness." Richard Feynman, in {\it Superstrings: A Theory of
Everything?}, edited by P. Davies and J. Brown (Cambridge University Press,
1988).

\bibitem{Donder} T. de Donder, {\it La Grafique Einsteinienne}
(Gauthier-Villars, Paris, 1921); V. A. Fock, {\it Theory of Space, Time and
Gravitation} (Pergamon, New York, 1959).

\bibitem{Eliezer} D. A. Eliezer and R. P. Woodard, Nucl. Phys. B{\bf 325}, 
389 (1989).

\bibitem{Susskind} L. Susskind, Phys. Rep. {\bf 104}, 181 (1984); E. 
Gildener, Phys. Rev. {\bf 14}, 1667 (1976); E. Gildner and S. Weinberg, 
Phys. Rev. {\bf 15}, 3333 (1976).

\bibitem{Capper} D. M. Capper, G. Leibrandt, and M. R. Medrano, Phys. Rev. 
{\bf 8}, 4320 (1973).

\bibitem{Veltman} G.`t Hooft and M. Veltman, Ann. Inst. Henri Poincar\'e, 
{\bf 30}, 69 (1974).
 
\bibitem{Veltman2} M. Veltman, {\it Reflections on the Higgs System}, 
Lectures given at CERN, 1996-1997, CERN 97-05 (1997).

\bibitem{Nath} P. Nath and M. Yamaguchi, hep-ph/9902323; hep-ph/9903298.

\bibitem{Graesser} M. L. Graesser, hep-ph/9902310; W. J. Marciano, 
hep-ph/9903451.

\bibitem{Pati} J. Pati and A. Salam, Phys. Rev. D{\bf 10}, 275 (1974).
 
\bibitem{Ross} G. G. Ross, {\it Grand Unified Theories}, Oxford University 
Press, 1984.
 
\bibitem{Gell-Mann2} M. Gell-Mann, P. Ramond, and R. Slansky, 
Rev. Mod. Phys. {\bf 50}, 721 (1978). 
 
\bibitem{Schmaltz} N. Arkani-Hamed and M. Schmaltz, hep-ph/9903417.

\bibitem{Hooft} G. 't Hooft, gr-qc/9903084 v3.

\end{thebibliography}
\end{document}